\documentclass[fleqn,twoside]{article}
\usepackage{espcrc2}

\usepackage{graphicx}
\usepackage[figuresright]{rotating}

\def\Journal#1#2#3#4{{#1} {#2}, #3 (#4)}

\def\NIMA{{Nucl. Instrum. Methods} A}

\def\PRL{Phys. Rev. Lett.}
\def\PRD{{Phys. Rev.} D}

\def\APJ{Astrophys. J.}
\def\SNP{Sov. Jour. Nucl. Phys.}

\newcommand{\AmS}{{\protect\the\textfont2
  A\kern-.1667em\lower.5ex\hbox{M}\kern-.125emS}}

\title{Solar Neutrino Precision Measurements using all 1496 Days
       of Super-Kamiokande-I Data}

\author{M. B. Smy\address[uci]{
        Department of Physics and Astronomy,\\
        University of California, Irvine,\\
        Irvine, California 92697, USA}
        (for the Super-Kamiokande Collaboration)}
       
\begin{document}

\begin{abstract}
The results of the entire Super-Kamiokande-I solar neutrino data are presented.
The measured interaction rate is $47\pm2\%$ of the rate expected
by the standard solar model and $133\pm5\%$ of the rate implied
by the SNO charged-current interaction rate. There is no evidence for
spectral distortion or a time dependent neutrino flux. Together
with the rates of other experiments, the Super-Kamiokande
results imply active solar
neutrino oscillations and restrict neutrino mixing and mass square
difference to lie within the LMA solution area.
\vspace{1pc}
\end{abstract}

\maketitle

\section{Introduction}
Between May 31$^{\mbox{st}}$, 1996 and
July 15$^{\mbox{th}}$, 2001 Super-Kamiokande (SK)
exposed 22,500 tons of purified water for
1496 days to solar neutrinos. The detector
is described elsewhere~\cite{linac}. Solar neutrinos
are detected using the Cherenkov
light emitted by the recoiling electron from
neutrino-electron elastic scattering. This
water Cherenkov detection technique offers
four advantages:
(i) the reconstructed direction of the recoiling electron
is strongly correlated with the neutrino direction,
(ii) the time of each neutrino interaction
is recorded, (iii) the neutrino spectrum can
be inferred from the
recoil electron spectrum,
and (iv) neutrino-elastic scattering is sensitive
to all neutrino flavors, although the cross section
for $\nu_\mu$ and $\nu_\tau$ is about six to seven times
less than for $\nu_e$.
However, only
high energy solar neutrinos can be observed in this
fashion; SK analyzes data above a threshold
of 5 MeV of the total energy
of the recoil electron. Only extremely few
solar neutrinos have an energy that high;
they arise from the $\beta^+$ decay of $^8$B
($^8$B neutrinos) and $^3$He-proton fusion
({\it hep} neutrinos) in the sun. The 
standard solar model (SSM) prediction
of the flux of those rare neutrinos
($\phi_{^8B}=5.05^{+1.01}_{-0.81}\times10^6/$cm$^2$s
and $\phi_{hep}=9.3\times10^3/$cm$^2$s with no uncertainty
given~\cite{ssm}) is
therefore difficult and beset by large
uncertainties. 

\begin{figure}[ht]
\vspace{9pt}
\includegraphics[width=7.5cm]{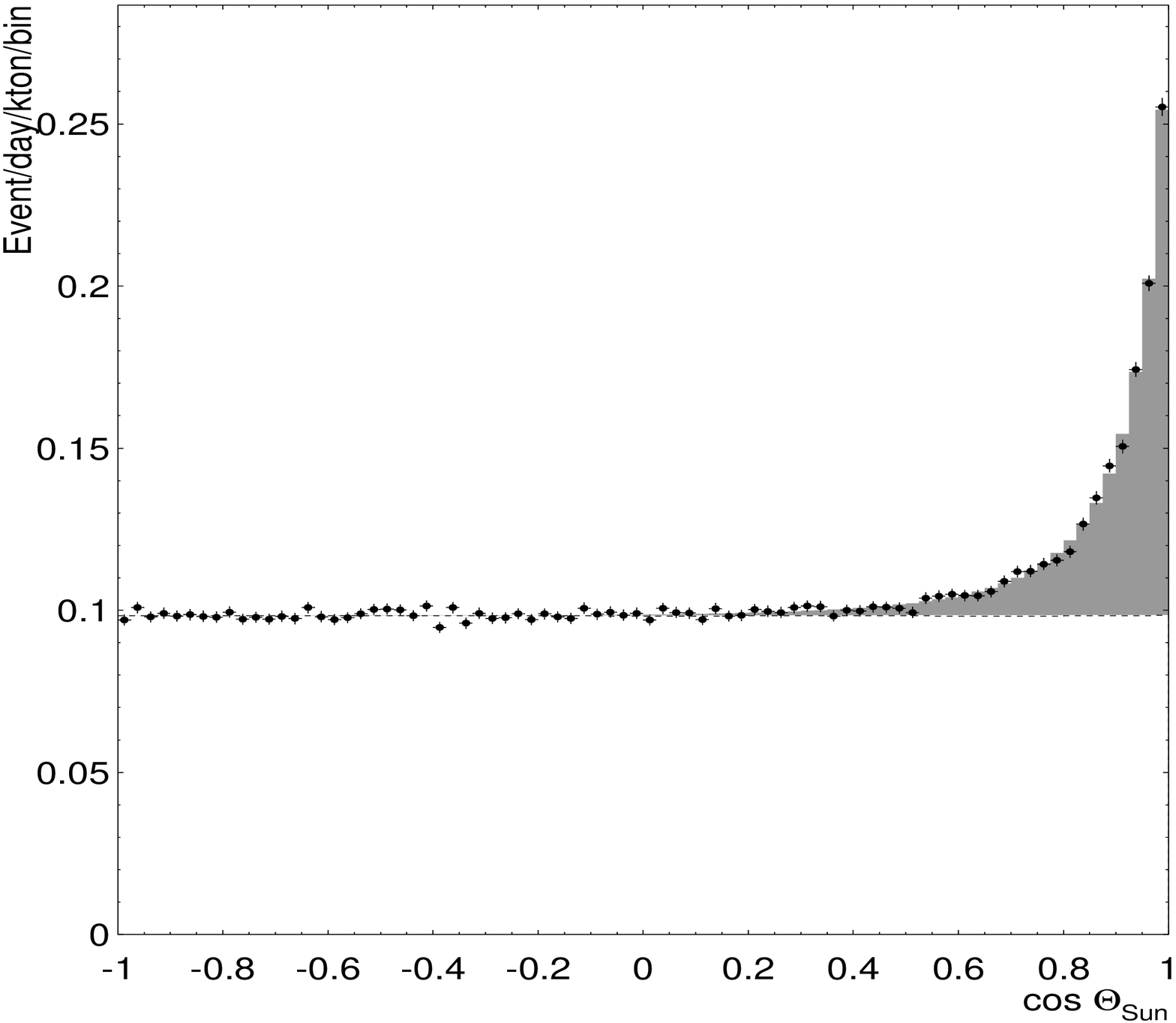}
\vspace*{-6.5cm}

\hspace*{1.2cm}\includegraphics[width=5cm]{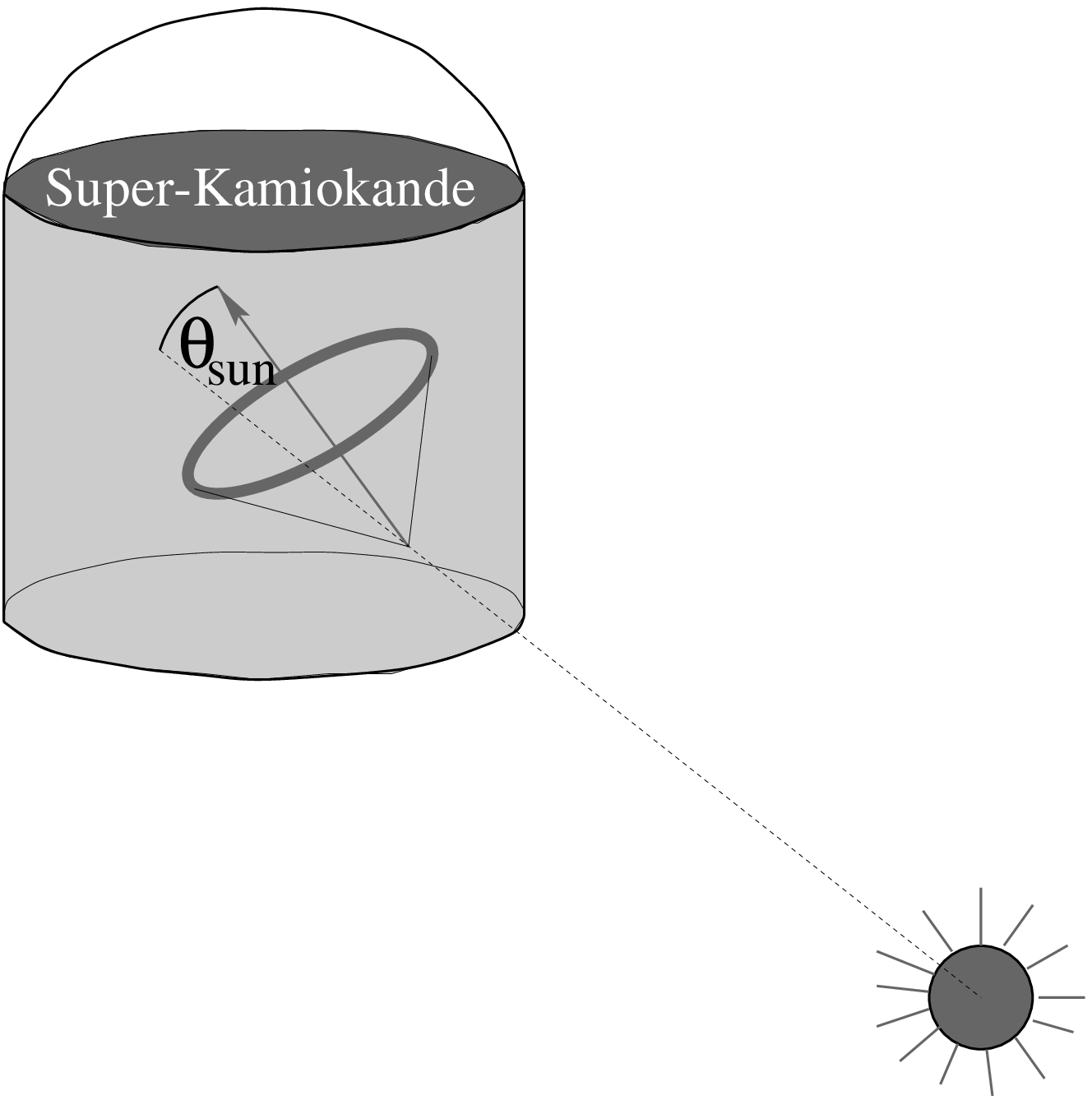}

\vspace*{1.0cm}
\caption{Angular Distribution of Solar Neutrino Event Candidates.}
\label{fig:cossun}
\end{figure}

\section{Solar Neutrino Flux}
\label{sec:flux}
SK uses the known position of the sun (at the time of
a solar neutrino event candidate) to statistically separate
solar neutrino interactions from background events.
During 1496 effective days 22,400$\pm800$
solar neutrino interactions were found in the gray shaded
forward peak of Figure~\ref{fig:cossun}.
From Monte Carlo calculations based on the SSM,
48,200$^{+9,600}_{-7,700}$ are expected. In spite of the large
neutrino flux uncertainty, the elastic scattering rate is significantly
less than expected from the SSM. The suppression factor 
Data/SSM~MC is about $3\sigma$ less than one:
0.465$\pm0.005$(stat.)$^{+0.016}_{-0.015}$(syst.)$^{+0.19}_{-0.17}$(SSM).
All solar neutrinos are born as $\nu_e$ in the sun;
if all solar neutrinos {\it detected by SK} are $\nu_e$'s as well
then a $\nu_e$ flux of
2.35$\pm0.02$(stat.)$\pm0.08$(syst.)$\times10^6/$cm$^2$s
leads to the
observed interaction rate.

SNO has
 measured the charged-current interaction rate
of solar $\nu_e$ with deuterium~\cite{sno}
(above a threshold neutrino energy of about 7 MeV) implying
a
$\nu_e$ flux of
1.76$^{+0.06}_{-0.05}$(stat.)$\pm0.10$(syst.) 
(assuming an undistorted $^8$B neutrino spectrum)
and consequently
16,800$^{+1,100}_{-1,000}$ $\nu_e$ interactions in SK.
SK has observed
1.334$\pm0.013$(stat.)$^{+0.047}_{-0.043}$(syst.)$^{+0.066}_{-0.061}$(SNO)
more events which is about 4.5$\sigma$ above one.

Beyond the $^8$B endpoint of $\approx15$ MeV, {\it hep} neutrinos
dominate the solar neutrino flux. Taking into account energy resolution,
an optimum {\it hep} neutrino search window (18 to 21 MeV)
was defined using Monte Carlo.
In that window $4.9\pm2.7$
events were observed; one {\it hep} neutrino interaction
is expected. This limits the {\it hep} flux to be less than
$73\times10^{3}/$cm$^2$s (or less than 7.9 times the SSM flux)
at 90\% C.L.

\begin{figure}[htb]
\vspace{9pt}
\includegraphics[width=7.5cm]{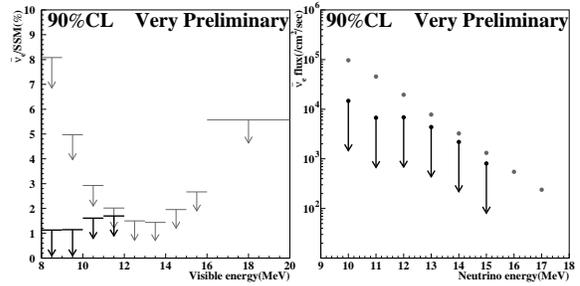}
\caption{Energy-Dependent Upper Limit of $\overline{\nu_e}$ at 90\% C.L.
The left panel assumes the $^8$B spectrum, the right
panel uses monochromatic neutrinos at eight energies.
The black data points statistically subtract spallation
background.}
\label{fig:nubarlimit}
\end{figure}

\section{Solar Antineutrinos}

If there are antineutrinos in the solar $\nu$ flux, they
could be detected by an inverse $\beta$ reaction
on hydrogen. Unlike elastic scattering (as shown in
Figure~\ref{fig:cossun}, those events
are reconstructed within about 60$^0$ of the
forward direction), the angular
distribution of this reaction is essentially flat,
with a slight backward bias, especially for low energy
neutrinos. The positron energy is strongly correlated
with the antineutrino energy.
After solar neutrino interactions are
removed using an angular cut of 60$^0$, the dominant
background at higher energy (above 8 MeV)
is due to cosmic ray-induced
spallation. Figure~\ref{fig:nubarlimit} shows two
upper flux limits at 90\% C.L.:
(i) all events below $\cos\theta_{\mbox{sun}}<0.5$
are considered to be antineutrino
interactions (ii) spallation background is statistically
subtracted. The limits of the left plot
(in \% of the SSM $^8$B flux)
assume antineutrinos with a $^8$B energy spectrum,
the limits given in the right plot (in /cm$^2$s)
are for
monochromatic antineutrinos of eight different
energies.

\begin{figure*}[bht]
\includegraphics[width=16.2cm]{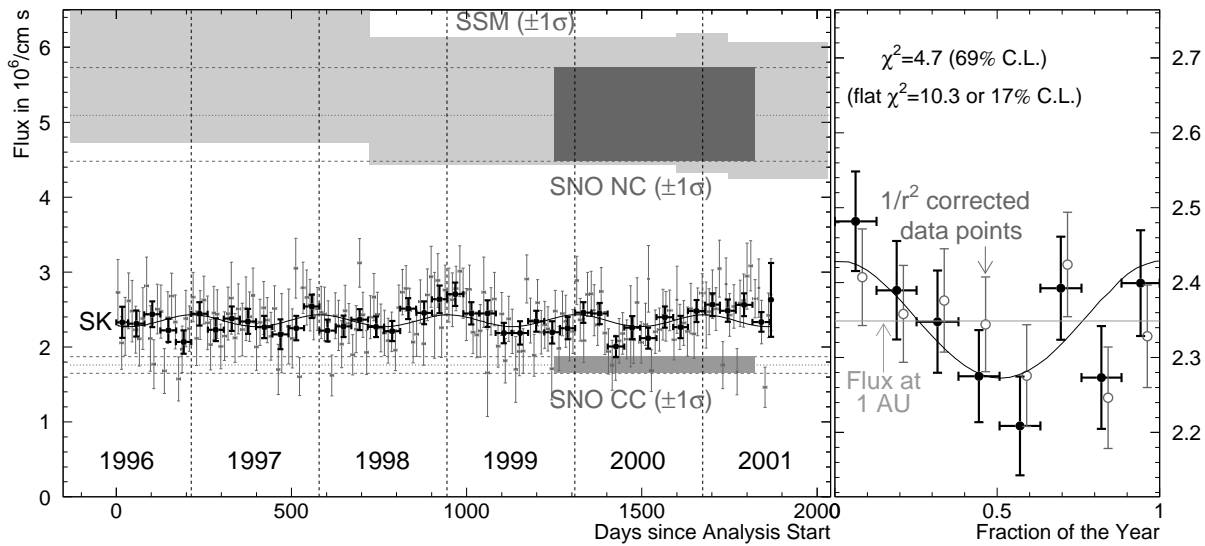}
\caption{Time Variation of the SK Elastic Scattering Flux. The gray
data points are measured every 10 days, the black data points every
1.5 months. The SSM prediction is indicated by the gray band; also
shown are the SNO charged-current and neutral-current measurements.
The black line indicates the expected annual 7\% flux variation.
The right-hand panel combines the 1.5 month bins to search for
yearly variations; the vertical scale is ten times larger and
the zero suppressed. The gray data points (open circles)
are obtained from the
black data points by subtracting the expected 7\% variation.}
\label{fig:timeseas}
\end{figure*}

\section{Time Variations}
Figure~\ref{fig:timeseas} shows the time variation of the
neutrino flux inferred from the SK rate in bins of ten
days and 1.5 months. A 7\% yearly variation is 
expected due to the 3.5\% change in distance between sun and
earth (assuming the neutrino flux is proportional to the
inverse square of this distance, that is, the sun is a neutrino
point source). The data favor such a
7\% variation over no variation by about 2.5$\sigma$.
No significant time variation is seen after subtraction of this effect.
To search for daily variation, the
data was binned according to solar zenith angle $\theta_z$.
Figure~\ref{fig:zenith} displays the resulting distribution.
No significant daily variation is found. The result can be summarized
by forming the ``day/night asymmetry'' $A_{DN}=\frac{D-N}{0.5(D+N)}$
(with $D$ meaning ``day rate'' and $N$ meaning ``night rate''):
$A_{DN}=-0.021\pm0.020$(stat.)$^{+0.013}_{-0.012}$(syst.) is
consistent with zero within 0.9$\sigma$.

\noindent\includegraphics[width=7.5cm]{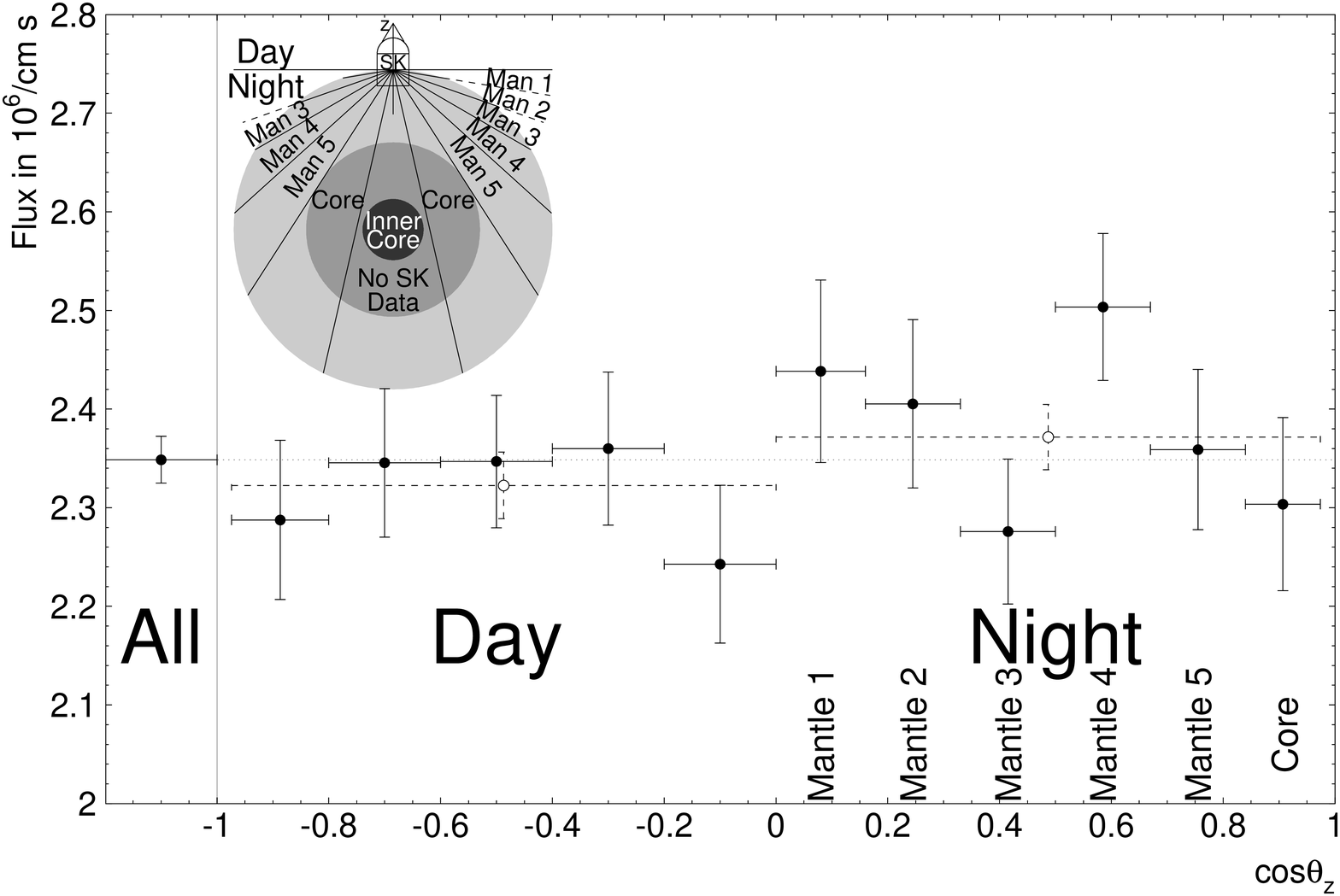}

\addtocounter{figure}{1}
\vspace{28pt}\noindent Figure \arabic{figure}.
Elastic Scattering Flux as a Function of Solar Zenith Angle.
Note the suppressed zero.
\label{fig:zenith}

\begin{figure}[htb]
\vspace{6pt}
\includegraphics[width=7.5cm]{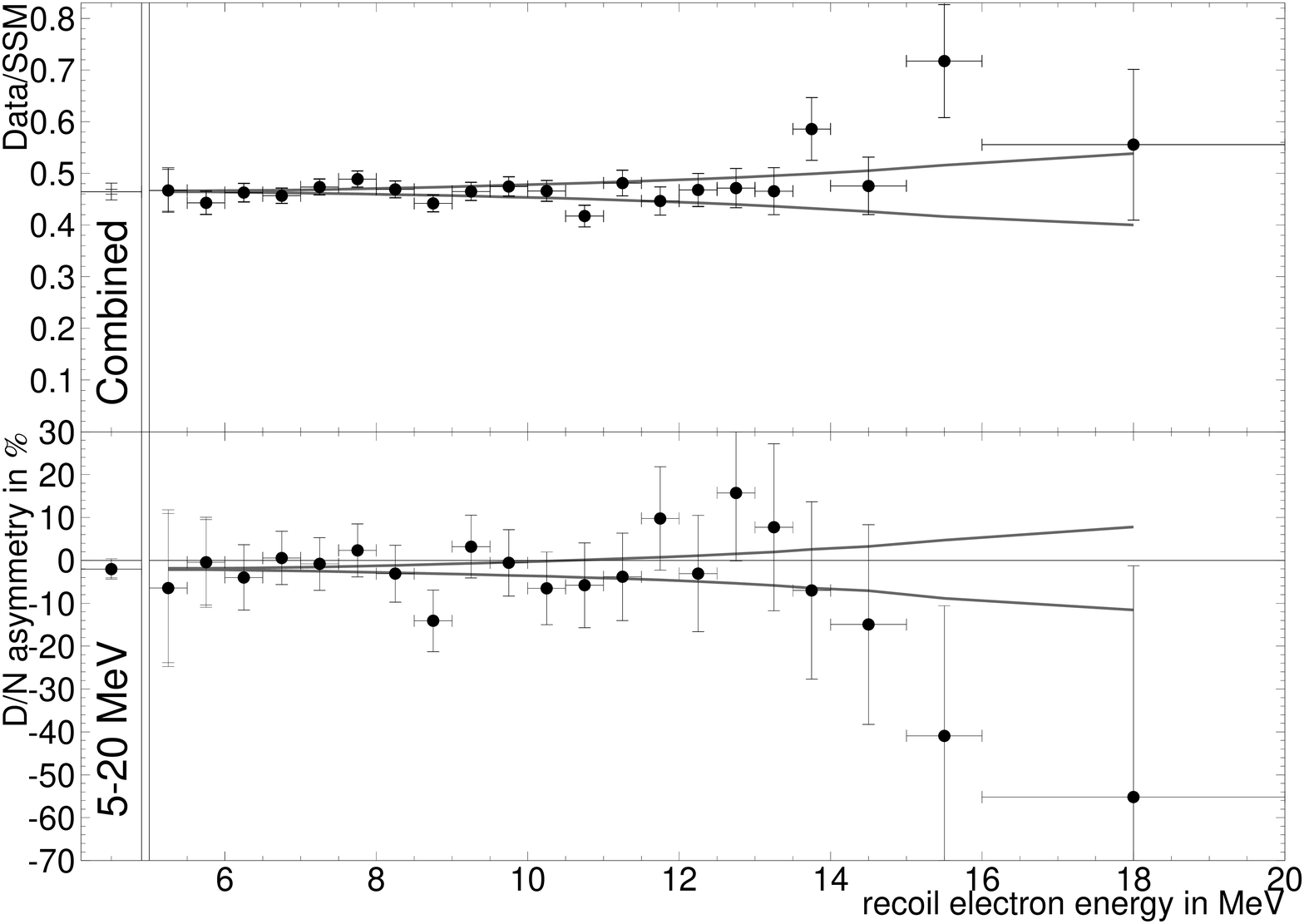}
\caption{Rate Suppression (Upper Panel)
and Day/Night Asymmetry (Lower Panel) as a Function
of Energy. The smaller error bars of each data point show the
statistical, the larger error bars the combined uncorrelated
uncertainty. The gray lines are the $\pm1\sigma$
correlated uncertainty due to the uncertainty in
the $^8$B $\nu$ spectrum
and the SK energy scale and resolution.
All data points are dominated by statistical uncertainty except
the combined rate.}
\label{fig:dnspec}
\end{figure}

\section{Spectrum}

For precision measurements of the spectrum it is necessary to
calibrate the absolute energy scale (and determine the energy resolution)
of SK with an outside source. We employed an electron linear accelerator
to determine the energy scale within 0.64\% and the resolution
within 2.5\%~\cite{linac}. In Figure~\ref{fig:dnspec} the observed
recoil electron spectrum is compared to the expected spectrum:
the ratio {\it Data/SSM} effectively plots spectral distortion. 
There is no significant spectral
distortion; the fit to the no-distortion hypothesis results
in $\chi^2=20.2/20$ d.o.f. corresponding to 44.3\% C.L.
The $\chi^2$ includes three systematic effects which will shift all
bins of the distribution in a correlated way (correlated uncertainties):
the uncertainty in (i) the $^8$B $\nu$ spectrum, (ii) the absolute
SK energy calibration, and (iii) the SK energy resolution.
The gray lines of Figure~\ref{fig:dnspec} show the size of the
sum (in quadrature) of (i), (ii) and (iii).
The same figure also plots the day/night asymmetry as a function
of recoil electron energy. The day/night asymmetry is consistent with
zero regardless of the recoil electron energy. 
The plotted correlated uncertainty is the sum of only
(ii) and (iii), since (i) is the same for the day and the night.

\section{Effects of Solar Neutrino Oscillations}

Solar neutrino oscillations explain both the
observed SK rate reduction compared to the SSM prediction
{\it and} the observed SK rate enhancement relative to
the SNO-based expectation (mentioned in section~\ref{sec:flux})
as conversion of $\nu_e$ into other active flavors, that
is $\nu_\mu$ or $\nu_\tau$. They also explain
the rates observed by various radio-chemical solar neutrino
experiments~\cite{cl,gallex,sage}. Solar neutrinos are
always born as $\nu_e$; their conversion probability into
$\nu_\mu$ or $\nu_\tau$ in vacuum is determined by their
energy and the distance of flight and the oscillation parameters.
For the purpose of this paper, we assume just two massive neutrinos;
in that case the oscillation parameters are the mixing angle
$\theta$ (controlling the relationship between flavor and
mass eigenstates) and the differences of the squared masses
between the two neutrinos $\Delta m^2$.

The presence of matter
strongly influences this conversion probability. In particular,
the matter density in the sun is high enough to induce
resonant conversion~\cite{msw} at solar neutrino energies
for a wide range of $\Delta m^2$ ($10^{-8}$ to $10^{-4}$eV$^2$).
In that case, the conversion probability is large or even maximal, even if the
mixing angle is small. The matter density in the earth also
affects the conversion probability leading to apparent daily
variations of the solar neutrino flux.

\begin{figure}[hbt]
\vspace{9pt}
\includegraphics[width=7.5cm]{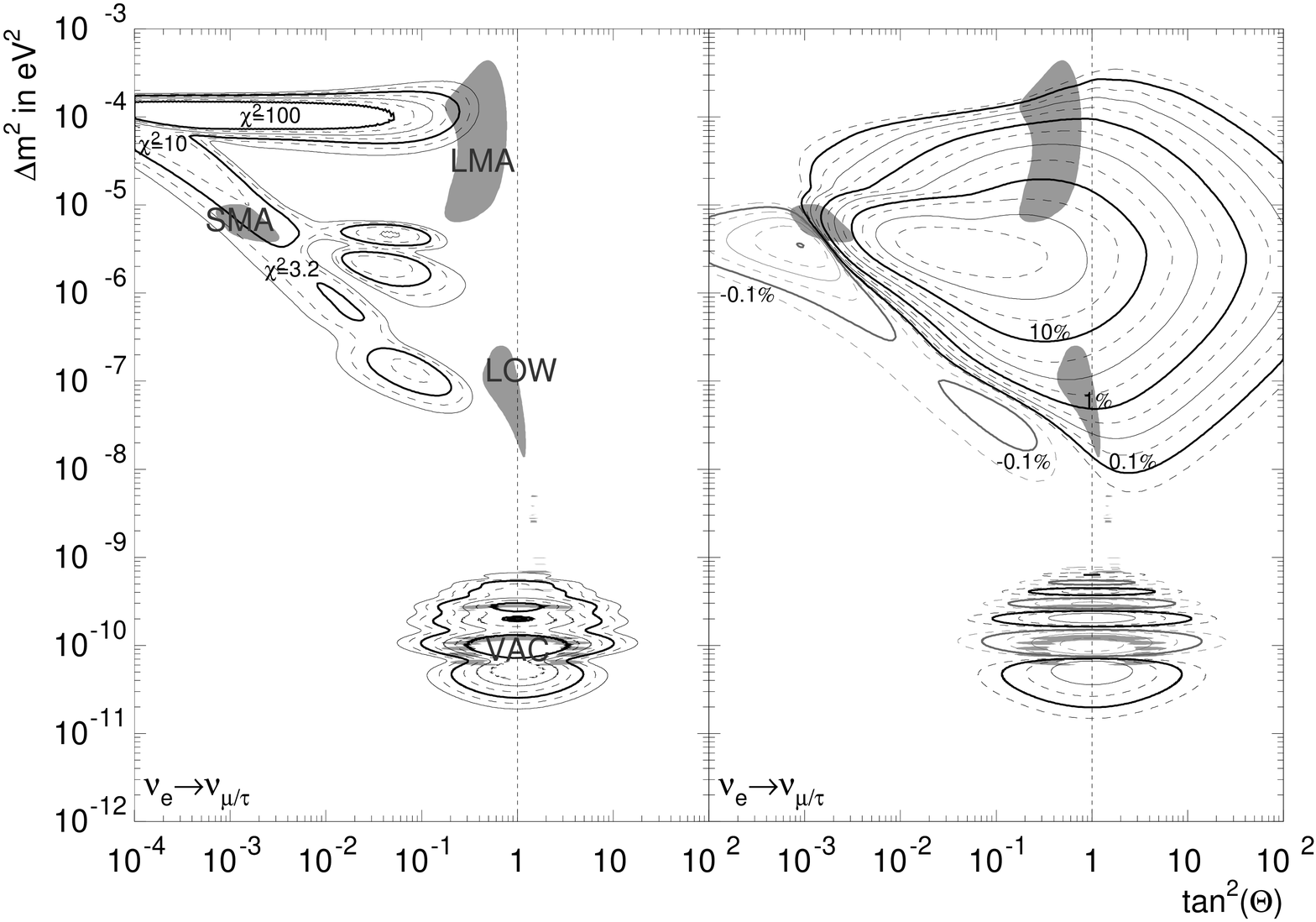}
\caption{Oscillation Predictions of SK
Spectral Distortion (Left) and Day/Night
Asymmetry (Right). The spectral distortion is
depicted in contours of equal $\chi^2$. The shown values of
both the $\chi^2$ and the day/night asymmetry contours
increase logarithmically.
Superimposed in gray are the usual solution areas.}
\label{fig:specdistdnexp}
\end{figure}

\begin{figure*}[bht]
\includegraphics[width=16.2cm]{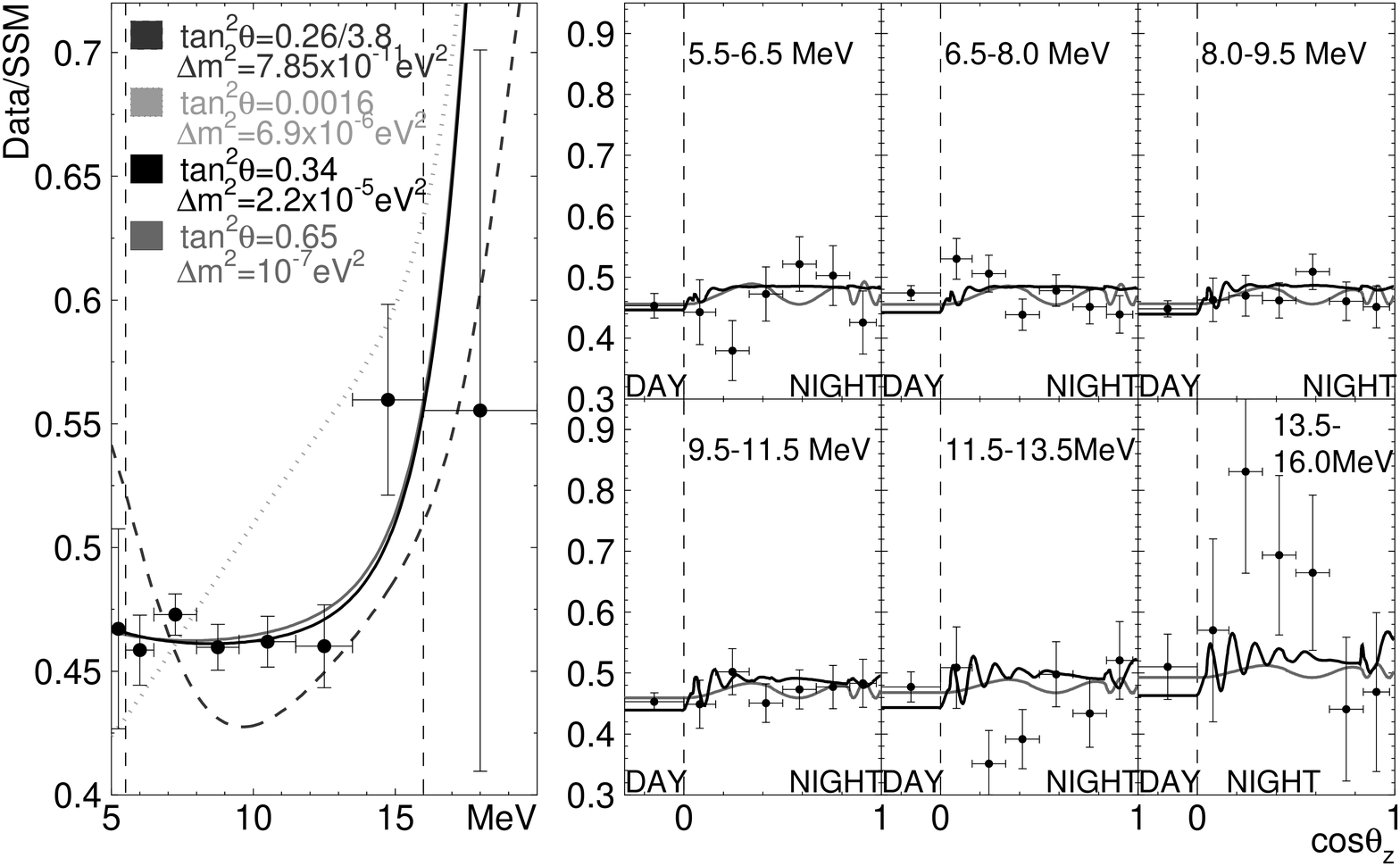}
\caption{Zenith Spectrum and Four Oscillation Predictions.
SMA (dotted line) and VAC (dashed line) solutions are disfavored
by spectral distortion data, LOW (light gray) solutions
by the absence of solar zenith angle variation. The LMA
solutions (black) fit best.}
\label{fig:zenspec}
\end{figure*}

Depending on $\theta$ and $\Delta m^2$, SK can therefore observe
distortions of the recoil electron spectrum, yearly variations
or daily variations. In Figure~\ref{fig:specdistdnexp}, the expected
spectral distortion and day/night asymmetry are shown for all
values of $\tan^2\theta$ and $\Delta m^2$ (The spectral distortion
is quantified by a $\chi^2$ method which uses the current SK
uncertainties as input). Overlaid are the traditional
oscillation ``solutions'', an area of parameter space capable of
explaining all measured solar neutrino rates.
The conversion probabilities for the
large mixing angle (LMA), small mixing angle (SMA)
and low $\Delta m^2$ (LOW) solutions are dominated by the matter
effects mentioned previously, while the vacuum solutions (VAC) match
the first oscillation phase maximum to the solar neutrino
energy range. LMA, LOW and VAC have a large mixing angle close to
$\pi/4$ while the mixing of the SMA is small (a few $10^{-2}$). 
SMA and VAC solutions predict a distorted spectrum, LMA and LOW
solutions daily variations.

\begin{figure*}[hbt]
\includegraphics[width=16.1cm]{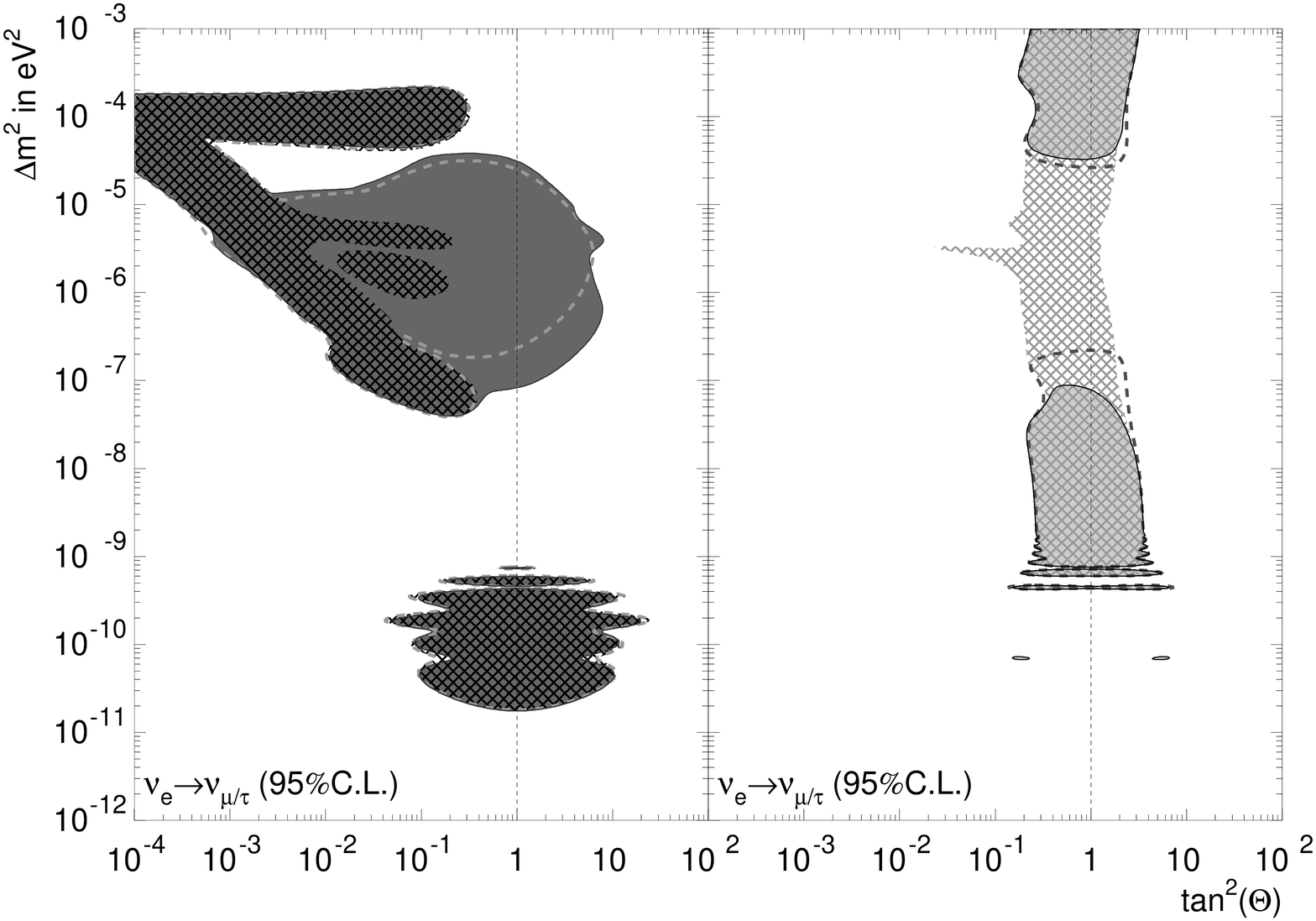}
\caption{SK Excluded (Left) and Allowed (Right) Regions
at 95\% C.L.
The cross hatched areas use the SK spectrum, the inside
of the dashed lines the shape of the SK day/night spectrum
(see Figure~\protect\ref{fig:dnspec}). The shaded areas
use the zenith angle spectrum. To get the allowed regions
on the right, the shape is combined with the SK rate and
the SSM constraint on the $^8$B flux.}
\label{fig:osc}
\end{figure*}

Figure~\ref{fig:zenspec} summarizes the data used for a
simultaneous search for spectral distortions and daily
variations: the zenith angle spectrum has eight energy bins
and seven solar zenith angle bins between 5.5 and 16 MeV. The $\chi^2$
for a flat shape hypothesis is $\chi^2=40.5/43$ d.o.f. which corresponds
to 58.1\% C.L. Overlaid are the oscillation predictions 
(fit to the data by varying freely the total $^8$B and {\it hep}
neutrino flux) for four
selected parameter pairs, each representing a solution.
The strong energy dependence of the flavor conversion
(inducing spectral distortions) predicted
by VAC and SMA is not supported by the data.
In case of LMA and LOW, there is little
energy dependence of the flavor conversion, the apparent spectral distortion
shown in Figure~\ref{fig:zenspec} is caused by a larger ratio of
the {\it hep} to the
$^8$B neutrino flux than predicted in the SSM. These two solutions
can only be distinguished by their daily flux variation predictions.
LMA shows rapid oscillations during the night. These oscillations
are washed out for the coarse bins of the data. The LOW solutions
however predict {\it slow} oscillations in the night.
The statistical fluctuations of the observed flat distributions
would be more consistent with slow oscillations of the opposite
phase.

\section{Oscillation Analysis}

Figure~\ref{fig:osc} shows the strong constraints
on solar neutrino oscillation parameters which are imposed by
the SK spectral distortion data (cross hatched areas)
and the SK zenith angle spectrum (gray areas).
A fit to two spectra (day and night; see gray dashed lines)
{\it partially} includes the zenith angle variation
which is {\it fully} included in the zenith angle spectrum
fit.
The analysis is described in~\cite{osc};
the definition of the $\chi^2$ employed to extract those
constraints can be found in~\cite{lma}. 
The areas in the left panel of Figure~\ref{fig:osc}
are excluded and use only the
(zenith angle) spectrum shape while the right panel
shows allowed areas based on this shape {\it and} the SK rate
(and the SSM $^8$B flux).
SK rate and spectrum allow only large
mixing angles. 
The large daily variation predicted between $\approx10^{-5}$
and $\approx10^{-7}$eV$^2$ splits this ``large mixing band'' into
two distinct allowed
areas (SK LMA and SK LOW/quasi-VAC).
The $\Delta m^2$ limits of these areas
are stronger for the zenith angle spectrum
compared to the day/night spectrum, especially
the upper limit for the SK LOW/quasi-VAC
which is reduced by a factor of two to three.

\vspace*{22pt}
\vspace*{9pt}
\noindent\includegraphics[width=7.5cm]{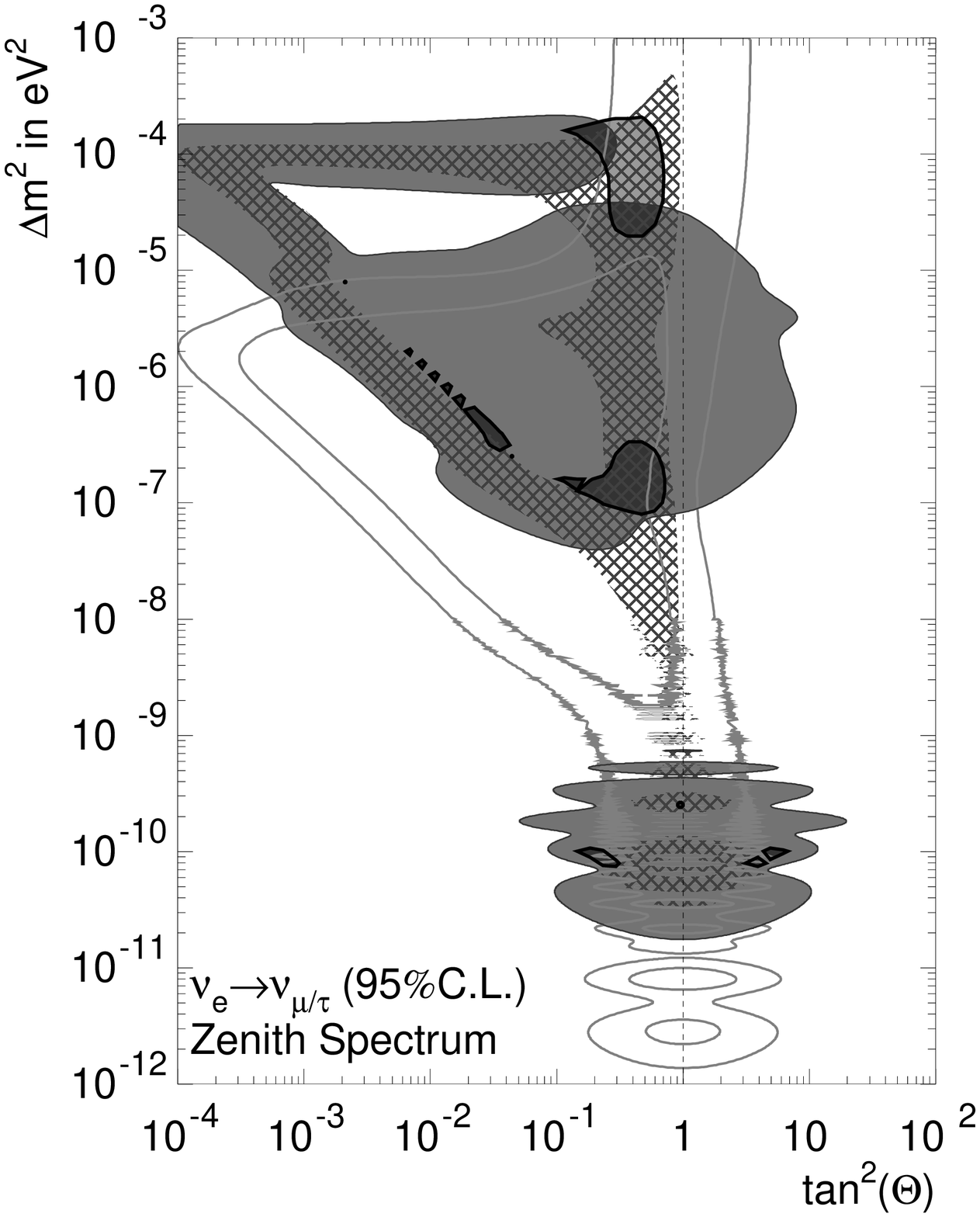}

\addtocounter{figure}{1}
\vspace{20pt}\noindent Figure \arabic{figure}.
\label{fig:zenspecosc1}
Area Excluded at 95\% C.L. by the
Shape of the SK Zenith Spectrum (Gray Area).
The exclusion is independent of all SSM $\nu$ flux
predictions. Overlaid is the allowed area
(inside thick black line shaded in light gray)
from the SNO day/night spectrum~\protect\cite{snodn}.
The
overlap of both is shaded dark gray. The cross hatched
area is allowed by the rate of the Homestake
experiment~\protect\cite{cl}
and the SSM prediction of that rate. Similarly, the
gray lines mark the boundary of the area allowed by
Gallex/GNO~\protect\cite{gallex} and
SAGE~\protect\cite{sage}.

Figure~9
compares the constraints of the zenith angle spectrum
(gray area) with
the allowed areas of the SNO day/night spectrum~\cite{snodn},
the Homestake experiment and the combined ``Gallium
experiments'' (Gallex/GNO and SAGE). All SNO small angle
($\tan^2\theta<0.1$; in particular the one
at $\tan^2\theta=0.002$, which is consistent with the
Gallium experiments), VAC (consistent
with Homestake and the Gallium experiments, including
the one at $\tan^2\theta=1$), and LOW solutions
(consistent with Homestake and the Gallium experiments)
are excluded. The SNO LMA solutions
(consistent with all experiments) are
restricted to a smaller range. The zenith angle spectrum
also excludes almost the entire Homestake allowed area
(except for LMA and the region just below the LOW
solutions).

\begin{figure}[hb]
\vspace{9pt}
\includegraphics[width=7.5cm]{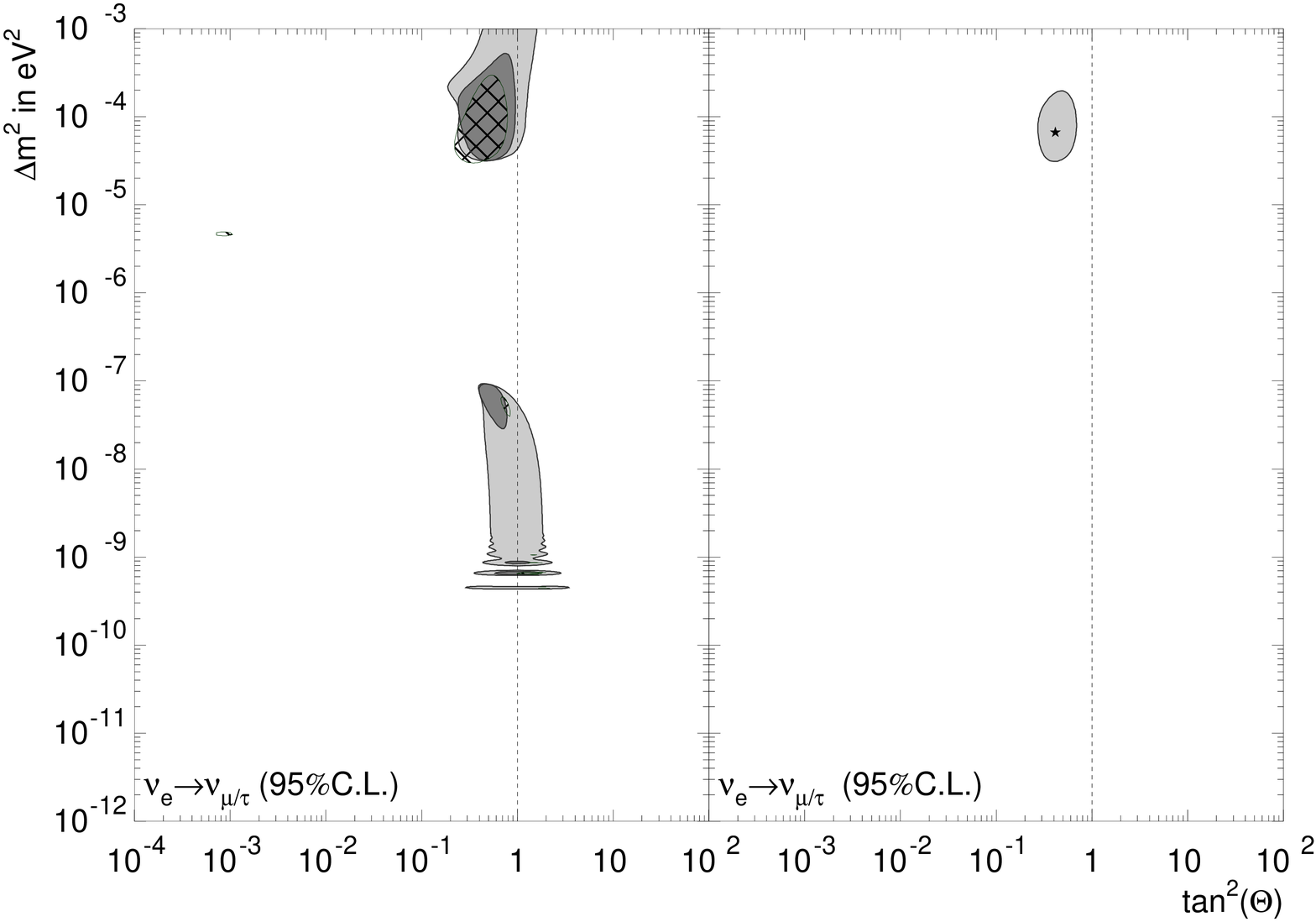}
\caption{Combined Allowed Regions.
The light gray area on the left is allowed
by the shape of the SK zenith angle spectrum combined
with the SK rate and the SNO charged-current rate.
The fit leading to the dark gray area includes in
addition the SNO neutral-current rate. The cross hatched
allowed area results from a combined fit of the SK zenith
angle spectrum and the Gallium, Homestake and SK rates
together (and the SSM $\nu$ flux predictions).
The right panel shows the results
of a fit to all solar neutrino data using
the SSM $\nu$ flux predictions except for the
$^8$B and {\it hep} fluxes.}
\label{fig:combined}
\end{figure}

\section{Combined Oscillation Fits}
Stronger constraints on neutrino oscillation parameters
can be obtained when the SK results are combined with
other solar neutrino data. In particular, for the first
time, a {\it unique} solution is obtained, when SK
results are combined with all other solar neutrino
data~\cite{lma}. The strongest constraints that are
independent of SSM neutrino flux predictions come from
fits which combine the SK zenith spectrum shape
with the SK and SNO rates~\cite{venice}.
The left plot of Figure~\ref{fig:combined} shows the
allowed regions of this fit. Superimposed
is a combined fit of the SK zenith spectrum shape and
the Gallium, Homestake and SK rates which relies on the SSM
neutrino flux predictions. In either case only large
mixing angle solutions survive. The LMA solutions are favored
in either case. When all solar neutrino data is combined,
the LMA solutions are more strongly favored.

\begin{figure}[bht]
\includegraphics[width=7.5cm]{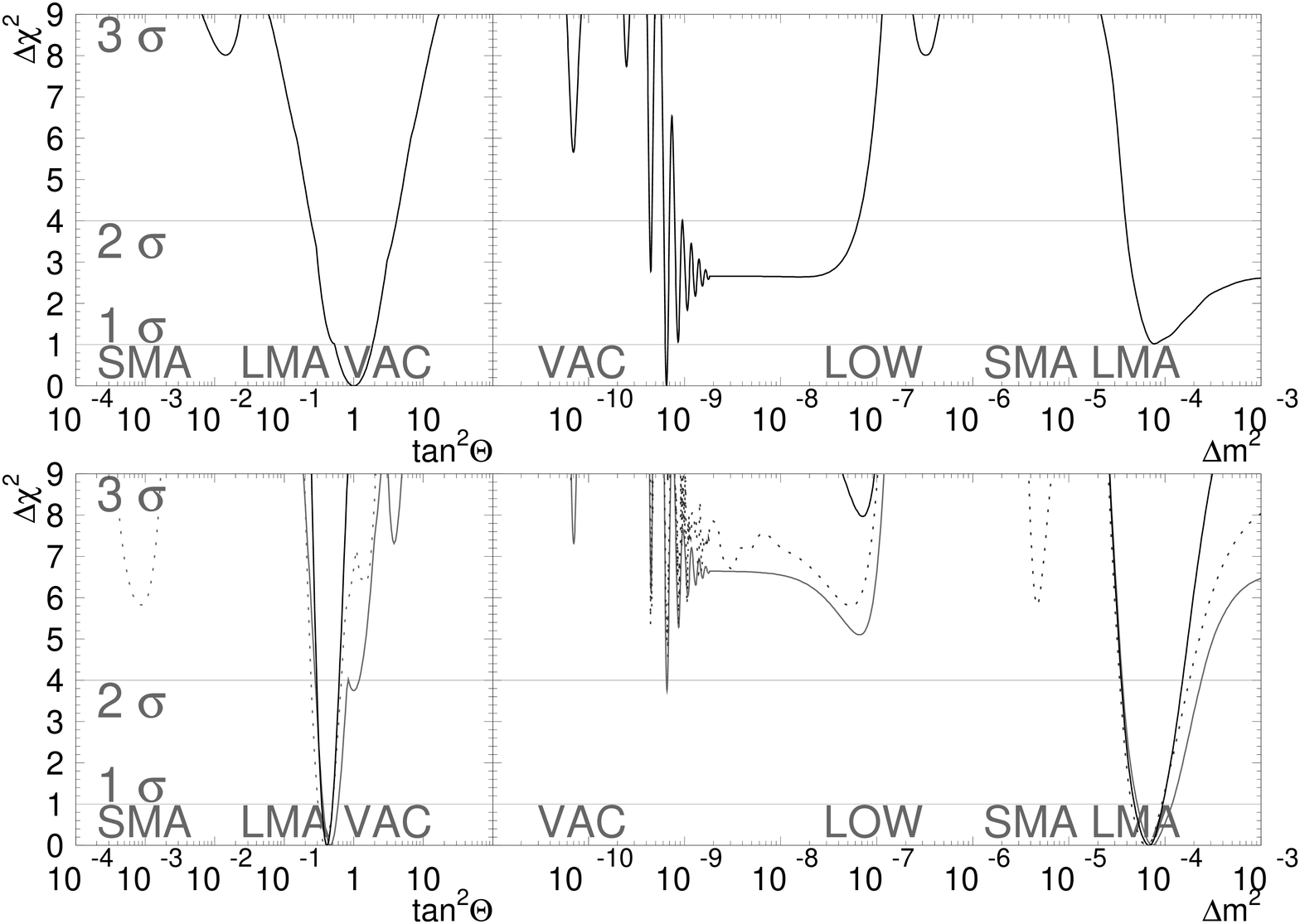}
\caption{Difference in $\chi^2$ as a Function of
$\tan^2\theta$ (left) and $\Delta m^2$ (right).
The top plots show only the fit to the SK
data and the SSM. The bottom
plots contain the fit to all solar data
(black solid line), the fit to SK and SNO
data (gray solid line) and the fit to Homestake,
Gallex/GNO, SAGE and SK data (and the SSM; gray
dotted line).}
\label{fig:osconed}
\end{figure}

Figure~\ref{fig:osconed} plots the $\chi^2$ difference
$\Delta\chi^2$ as a function of only one parameter
for various fits. (The
other parameter is chosen as to minimize $\chi^2$.)
$\Delta\chi^2$
has the statistics of a $\chi^2$ with one
degree of freedom.
SK data by itself strongly disfavors small mixing.
LMA solutions are favored, however quasi-VAC
solutions (between LOW and VAC)
provide reasonable fits as well with small ``pockets''
of parameter space
that fit even better than LMA solutions. The SK-SNO
combined fit favors the LMA at the 2$\sigma$ level
and tightens the allowed range in the mixing angle.
The SK-Gallium-Homestake combined fit disfavors
LOW and quasi-VAC solutions more strongly, but SMA
solutions are only disfavored by 2 to 2.5$\sigma$.
The all-combined fit favors the LMA by almost
3$\sigma$ with small allowed ranges in
$\tan^2\theta$ and $\Delta m^2$.

\section{Conclusion}
Super-Kamiokande solar neutrino precision measurements
provide the first hint of solar neutrino flavor
conversion due to oscillation in a {\it unique}
parameter region, the LMA solution.
Solar neutrino oscillation parameters that lead
to significant distortions of the $^8$B neutrino spectrum
or solar zenith angle variations are excluded by Super-Kamiokande
data.

\end{document}